\documentclass[journal]{IEEEtran}
\usepackage{amsmath,epsfig,amssymb,verbatim,amsopn,citesort,subfigure,color,multirow}

\newcommand{\ve}[1]{\boldsymbol{#1}}

 \newcommand{\vEe}{\ve{e}}
 
\newcommand{\vG}{\ve{G}} \newcommand{\vg}{\ve{g}}
 \newcommand{\vh}{\ve{h}}
\newcommand{\vI}{\ve{I}}

 \newcommand{\vv}{\ve{v}}
\newcommand{\vW}{\ve{W}} \newcommand{\vw}{\ve{w}}
 \newcommand{\vx}{\ve{x}}
 \newcommand{\vy}{\ve{y}}

\newcommand{\Na}{N_{A}}

\newcommand{\Ne}{N_{E}}

\newcommand{\con}[1]{{#1}^{\ast}}
\newcommand{\mct}[1]{{#1}^{\dagger}}

\newcommand{\su}{\sigma^2_{u}}
\newcommand{\sv}{\sigma^2_{v}}

\newcommand{\sh}{\sigma^2_{h}}
\newcommand{\shtilde}{\sigma^2_{\tilde{h}}}
\newcommand{\shhat}{\sigma^2_{\hat{h}}}

\def\onedot{.,\,}

\def\eg{e.g\onedot} 

\def\ie{i.e\onedot}

\def\wrt{w.r.t.\,}

\newcounter{mytempeqncnt}

\newcommand{\AuthorOne}{Xiangyun Zhou}
\newcommand{\AuthorTwo}{Matthew R. McKay}

\newcommand{\ThankOne}{This paper was presented in part at the Int. Conf. on Signal Processing and Commun. Syst., Omaha, NE, Sept.
2009.}

\newcommand{\ThankTwo}{Xiangyun Zhou is with the Research School of Information Sciences and Engineering,
the Australian National University, Canberra, ACT 0200, Australia.
(Email: xiangyun.zhou@anu.edu.au). His work was supported under Australian Research Council's Discovery Projects funding scheme (project no. DP0773898). Matthew R. McKay is with the
Department of Electronic and Computer Engineering, Hong
Kong University of Science and Technology, Hong Kong.
(Email: eemckay@ust.hk).}

\newcommand{\ThankThree}{The authors would like to thank Dr. Parastoo Sadeghi for useful discussions.}

\begin{document}

\title{Secure
Transmission with Artificial Noise over Fading Channels: Achievable
Rate and Optimal Power Allocation}

\author{
\authorblockN{\AuthorOne, \:\textit{Student Member, IEEE},\:\,and\:\,\AuthorTwo, \:\textit{Member, IEEE}
\thanks{\ThankOne}
\thanks{\ThankTwo}
}}

\maketitle

\pagestyle{empty}

\begin{abstract}

We consider the problem of secure communication with multi-antenna
transmission in fading channels. The transmitter simultaneously
transmits an information bearing signal to the intended receiver and
artificial noise to the eavesdroppers. We obtain an analytical
closed-form expression of an achievable secrecy rate, and use it as
the objective function to optimize the transmit power allocation
between the information signal and the artificial noise. Our
analytical and numerical results show that equal power allocation is
a simple yet near optimal strategy for the case of non-colluding
eavesdroppers. When the number of colluding eavesdroppers increases,
more power should be used to generate the artificial noise. We also
provide an upper bound on the signal-to-noise ratio (SNR) above
which the achievable secrecy rate is positive and show that the
bound is tight at low SNR. Furthermore, we consider the impact of
imperfect channel state information (CSI) at both the transmitter
and the receiver and find that it is wise to create more artificial
noise to confuse the eavesdroppers than to increase the signal
strength for the intended receiver if the CSI is not accurately
obtained.

\end{abstract}


\section{Introduction} \label{sec:intro}

Security is a fundamental problem in wireless communications due to
the broadcast nature of the wireless medium. Traditionally, secure
communication is achieved by using cryptographic technologies such
as encryption. On the other hand, the studies from an
information-theoretic viewpoint have found conditions for reliable
secure communication without using secret keys. In the pioneering
works on information-theoretic security, Wyner introduced the
wiretap channel model in which the eavesdropper's channel is a
degraded version of the receiver's channel~\cite{wyner_75}.
Csisz\'{a}r and K\"{o}rner considered a general non-degraded channel
condition and studied the transmission of both a common message to
two receivers and a confidential message to only one of
them~\cite{csiszar_78}. The results in these early works showed that
a positive secrecy capacity can be achieved if the intended receiver
has a better channel than the eavesdropper.

Recently, information-theoretic security with multi-antenna
transmission has drawn a lot of attention. Many works have been
devoted to analyzing the secrecy capacity with various antenna
configurations and channel conditions,
\eg~\cite{li_07,shafiee_07,khisti_09b,oggier_08}. With multiple
antennas at the transmitter, the optimal input structure (for
Gaussian codes) that maximizes the secrecy rate of Gaussian
channels was found to be in the form of beamforming
transmission~\cite{li_07,shafiee_07}. The secrecy capacity of
Gaussian channels with multiple antennas at both the transmitter and
the receiver was obtained in~\cite{khisti_09b,oggier_08}. One of the
main assumptions in the above-mentioned works is that the
eavesdropper's channel is known at the transmitter. Clearly this
assumption is usually impractical, especially for fading channels.
The ergodic secrecy capacity with and without knowing the
eavesdropper's channel was studied for fading channels
in~\cite{parada_05,barros_06,liang_08,gopala_08,khisti_09a}. The authors in~\cite{liang_08} studied a fading broadcast channel with confidential information intended only for one receiver and derived the optimal power allocation that minimizes the secrecy outage probability. The authors in~\cite{gopala_08} proposed an on-off power transmission with variable-rate allocation scheme for single antenna systems, which was shown to approach the optimal performance at asymptotically high signal-to-noise ratio (SNR). The authors in~\cite{khisti_09a} extended the ergodic secrecy capacity result to systems with multiple antennas and developed capacity bounds in the large antenna limit.

Furthermore, various physical-layer techniques were proposed to
achieve secure communication even if the receiver's channel is worse
than the eavesdropper's channel. One of the main techniques is the
use of interference or artificial noise to confuse the eavesdropper.
With two base stations connected by a high capacity backbone, one
base station can simultaneously transmit an interfering signal to
secure the uplink communication for the other base
station~\cite{jorgensen_07,simeone_08}. In the scenario where the
transmitter has a helping interferer or a relay node, the secrecy
level can also be increased by having the interferer~\cite{tang_08}
or relay~\cite{lai_08} to send codewords independent of the source
message at an appropriate rate. When multiple cooperative nodes are
available to help the transmitter, the optimal weights of the signal
transmitted from cooperative nodes, which maximize an achievable
secrecy rate, were derived for both
decode-and-forward~\cite{dong_08} and
amplify-and-forward~\cite{dong_09} protocols. The use of
interference for secrecy is also extended to multiple-access and
broadcast channels with user
cooperation~\cite{tekin_08,ekrem_08,ulukus_09}.

When multiple antennas are available at the transmitter, it is
possible to simultaneously transmit both the information bearing
signal and artificial noise to achieve secrecy in a fading
environment~\cite{negi_05,goel_08,swindlehurst_09}. The artificial
noise is radiated isotropically to mask the transmission of the
information signal to the intended receiver. In the design of this
multi-antenna technique, the transmit power allocation between the
information signal and the artificial noise is an important
parameter, which has not been investigated
in~\cite{negi_05,goel_08}. A sub-optimal power allocation strategy
was considered in~\cite{swindlehurst_09}, which aims to meet a
target signal to interference and noise ratio at the intended
receiver to satisfy a quality of service requirement.

In this paper, we study the problem of secure communication in
fading channels with a multi-antenna transmitter capable of
simultaneous transmission of both the information signal and the
artificial noise. We derive a closed-form expression for an
achievable secrecy rate in fading channels. The availability of a
closed-form secrecy rate expression greatly reduces the complexity
of obtaining the optimal power allocation between transmission of
the information signal and the artificial noise. We also study the
critical SNR above which the achievable secrecy rate is positive.
This is an important problem in wideband communications in which a
higher throughput is achieved by reducing the SNR per hertz while
increasing the bandwidth~\cite{verdu_02}. Furthermore, perfect
channel state information (CSI) at both the transmitter and the
receiver is usually assumed in the existing studies on
information-theoretic security. With this assumption, the artificial
noise is accurately transmitted into the null space of the intended
receiver's channel. When the CSI is not perfectly known at the
transmitter, the artificial noise leaks into the receiver's channel.
The effects of imperfect CSI on the achievable secrecy rate and the
aforementioned design parameters are investigated in this paper.

The main contributions of this work are summarized as follows:

\begin{itemize}

\item

In Section~\ref{sec:SecCap}, we derive analytical closed-form lower
bounds on the ergodic secrecy capacity for both non-colluding and
colluding eavesdroppers. These closed-form expressions, which give
achievable secrecy rates for secure communications with artificial
noise, greatly reduce the complexity of system design and analysis,
and also allow analytical insights to be obtained.

\item
In Section~\ref{sec:PowAll}, we study the optimal power allocation
between transmission of the information signal and the artificial
noise. For the non-colluding eavesdropper case, the equal power
allocation is shown to be a simple strategy that achieves nearly the
same secrecy rate as the optimal power allocation. For the colluding
eavesdropper case, more power should be used to transmit the
artificial noise as the number of eavesdropper increases. Analytical
results are obtained in the high SNR regime in both cases.

\item
In Section~\ref{sec:CriSNR}, we derive an upper bound on the
critical SNR above which the achievable secrecy rate is positive.
The bound is shown to be tight at low SNR, hence is useful in the
design and analysis of wideband secure communications.

\item
In Section~\ref{sec:ImpCSI}, we derive an ergodic secrecy capacity
lower bound taking into account channel estimation errors and
investigate the effects of imperfect CSI on the optimal power
allocation and the critical SNR for secure communication. In
particular, we find that it is better to create more artificial
noise for the eavesdroppers than to increase the signal strength for
the intended receiver as the channel estimation error increases.

\end{itemize}

Throughout the paper, the following notations will be used: Boldface
upper and lower cases denote matrices and vectors, respectively.
$[\cdot]^T$ denotes the matrix transpose operation,
$\con{[\cdot]}$ denotes the complex conjugate operation, and
$\mct{[\cdot]}$ denotes the conjugate transpose operation. The
notation $E\{\cdot\}$ denotes the mathematical expectation.
$\|\cdot\|$ denotes the norm of a vector
and $|\cdot|$ denotes the determinant of a matrix.

\section{System Model} \label{sec:SysMod}

We consider secure communication between a transmitter (Alice) and a
receiver (Bob) in the presence of eavesdroppers (Eves). Alice has
$\Na$ antennas ($\Na > 1$) and Bob has a single antenna. This
scenario is representative, for example, of downlink transmission in
cellular systems and wireless local area networks. In addition, each
Eve is equipped with a single antenna. We consider two cases, namely
non-colluding and colluding eavesdroppers. In the former case, Eves
individually overhear the communication between Alice and Bob
without any centralized processing. While in the latter case, there
are $\Ne$ Eves capable of jointly processing their received
information. Therefore, the non-colluding case can be seen as a
special colluding case where $\Ne=1$. We assume that $\Na
> \Ne$ for which the reason will become clear in the next section.
We also assume that Eves are passive, hence they cannot transmit
jamming signals. The received symbols at Bob and the multiple
colluding Eves are given by, respectively,
\begin{eqnarray}
    y_B &=& \vh \vx + n,\label{eq:SigMod1a}\\
    \vy_E &=& \vG \vx + \vEe,\label{eq:SigMod2a}
\end{eqnarray}
where $\vh$ is a $1 \times \Na$ vector denoting the channel between
Alice and Bob and $\vG$ is an $\Ne \times \Na$ matrix denoting the
channel between Alice and multiple colluding Eves. The elements of
$\vh$ and $\vG$ are independent zero-mean complex Gaussian random
variables. $n$ and $\vEe$ are the additive white Gaussian noises at
Bob and Eves, respectively. Without loss of generality, we normalize
the variance of $n$ to unity. We assume that $\vh$ is accurately
estimated by Bob and is also known by Alice using a noiseless
feedback link from Bob\footnote{A reliable feedback link could be
achieved by using low rate transmission with appropriate
quantization schemes. The design of a high-quality feedback link and
the effect of noisy feedback is beyond the scope of this work.
However, we will investigate the effect of imperfect channel
knowledge at Alice by considering channel estimation errors at Bob
in Section~\ref{sec:ImpCSI}.}. Similar to~\cite{negi_05}, we assume
that the knowledge of both $\vh$ and $\vG$ is available at Eve,
which makes the secrecy of communication independent of the secrecy
of channel gains.

The key idea of guaranteeing secure communication using artificial
noise proposed in~\cite{negi_05} is outlined as follows. We let an
$\Na \times \Na$ matrix $\vW = [\vw_1 \,\,\, \vW_2]$ be an
orthonormal basis of $\mathbb{C}^{\Na}$, where $\vw_1 =
\mct{\vh}/\|\vh\|$. The $\Na \times 1$ transmitted symbol vector at
Alice is given by $\vx = \vw_1 u + \vW_2\vv$, where the variance of
$u$ is $\su$ and the $\Na-1$ elements of $\vv$ are independent and
identically distributed (i.i.d.) complex Gaussian random variables
each with variance $\sv$. $u$ represents the information bearing
signal and $\vv$ represents the artificial noise. The received
symbols at Bob and Eves become
\begin{eqnarray}
    y_B &=& \vh \vw_1 u + \vh \vW_2 \vv + n = \|\vh\| u +
    n,\label{eq:SigMod1}\\
    \vy_E &=& \vG \vw_1 u + \vG \vW_2 \vv + \vEe
    = \vg_1 u + \vG_2 \vv + \vEe,\label{eq:SigMod2}
\end{eqnarray}
where we have defined that $\vg_1 = \vG \vw_1$ and $\vG_2 =  \vG
\vW_2$.

We consider a total power per transmission denoted by $P$, that is,
$P = \su + (\Na-1)\sv$. Due to the normalization of the noise
variance at Bob, we also refer to $P$ as the transmit
SNR. One important design parameter is the ratio of power
allocated to the information bearing signal and the artificial
noise. We denote the fraction of total power allocated to the
information signal as~$\phi$. Hence, we have the following
relationships:
\begin{eqnarray}
    \su &=& \phi P, \label{eq:PowerFactor1}\\
    \sv &=& (1-\phi)P/(\Na-1).\label{eq:PowerFactor2}
\end{eqnarray}

Since $\vh$ is known by Alice, she can adaptively change the
value of $\phi$ according to the instantaneous realization of $\vh$.
We refer to this strategy as the adaptive power allocation strategy.
Alternatively, Alice can choose a fixed value for $\phi$
regardless of the instantaneous channel realization, which we refer
to as the non-adaptive power allocation strategy. Note that Alice
does not know $\vG$, and thus equally distributes the transmit power
amongst the artificial noise signal, as given by (\ref{eq:PowerFactor2}).

\section{Secrecy Capacity Lower Bound} \label{sec:SecCap}

The secrecy capacity is the maximum transmission rate at which the
intended receiver can decode the data with arbitrarily small error
while the mutual information between the transmitted message and the
received signal at the eavesdropper is arbitrarily small. It is
bounded from below by the difference in the capacity of the channel
between Alice and Bob and that between Alice and
Eve~\cite{csiszar_78}. In this section, we derive a closed-form
expression for an ergodic secrecy capacity lower bound with
transmission of artificial noise.

The capacity of the channel between Alice and Bob is given by
\begin{eqnarray}\label{eq:AveCap1}
    C_1 &=& E_{\vh} \{\log_2 (1 + \su \|\vh\|^2)\}\nonumber\\
    &=& E_{\vh} \{\log_2 (1 + \phi P \|\vh\|^2)\}.
\end{eqnarray}

Without loss of generality, we normalize the variance of each
element of $\vh$ to unity. It is then easy to see that $\|\vh\|^2$
follows a Gamma distribution with parameters $(\Na,1)$. Therefore,
for systems with non-adaptive power allocation strategy, we can
rewrite (\ref{eq:AveCap1}) in an integral form as
\begin{eqnarray}
    C_1 &=& \frac{1}{\ln2}\int_0^{\infty} \ln(1+\phi P x) x^{\Na-1}
    \frac{\exp(-x)}{\Gamma(\Na)}\mathrm{d}x,\nonumber\\
    &=& \frac{1}{\ln2}\exp\Big(\frac{z}{P}\Big)
    \sum_{k=1}^{\Na}\text{E}_{k}\Big(\frac{z}{P}\Big),\label{eq:AveCap1C}
\end{eqnarray}
where $\Gamma(\cdot)$ is the Gamma function, $\text{E}_n(\cdot)$ is
the generalized exponential integral, (\ref{eq:AveCap1C}) is
obtained using an integral identity given in~\cite{alfano_04}, and
we have defined $z = \phi^{-1}$.

Next we study the capacity of the channel between Alice and multiple
colluding Eves. When multiple Eves are present, the noise at each
Eve may be different. In addition, the receiver noise levels at Eves
may not be known by Alice and Bob. To guarantee secure
communication, it is therefore reasonable to consider the worst case
scenario where the noises at Eves are arbitrarily small. Note that
this approach was also taken in~\cite{goel_08}. In this case, we can
normalize the distance of each Eve to make the variance of the
elements of $\vG$ equal to unity without loss of
generality.\footnote{With the noiseless eavesdropper assumption, the
capacity between Alice and each Eve is determined from the
signal-to-artificial-noise ratio. Considering the signal reception
at a particular Eve, both the information signal and the artificial
noise are generated from the same source (Alice), and hence their
ratio is independent of the large scale fading from Alice to Eve.
That is to say, the signal-to-artificial-noise ratios are i.i.d.
random variables for all Eves, regardless of their distances from
Alice.}

The noiseless eavesdropper assumption effectively gives an upper
bound on the capacity of the channel between Alice and multiple
colluding Eves as
\begin{eqnarray}\label{eq:Cap2}
    \!\!\!\!C_2 \!\!\!\!&=&\!\!\!\! E_{\vh, \vg_1, \vG_2}\Big\{\log_2\Big|\vI+\su\vg_1
    \mct{\vg_1} (\sv\vG_2\mct{\vG_2})^{-1}\Big|\Big\}\nonumber\\
     \!\!\!&=&\!\!\! E_{\vh, \vg_1, \vG_2}\Big\{\!\log_2\Big(1\!+\!\frac{\Na\!-\!1}{z\!-\!1}
    \mct{\vg_1} (\vG_2\mct{\vG_2})^{-1}\vg_1\Big)\!\Big\},
\end{eqnarray}
where we have again used $z = \phi^{-1}$. The expectation over $\vh$
in (\ref{eq:Cap2}) is due to the fact that $z$ may be dependent on
$\vh$ (which happens when adaptive power allocation strategy is
used). It is required in (\ref{eq:Cap2}) that $\vG_2\mct{\vG_2}$ is
invertible, which is guaranteed with the assumption of $\Na > \Ne$.
If the assumption is violated, the colluding eavesdroppers are able
to eliminate the artificial noise, resulting $C_2 = \infty$. Hence,
We assume $\Na > \Ne$ for guaranteeing secure communication.

Since $\vG$ has i.i.d. complex Gaussian entries and $\vW$ is a
unitary matrix, $\vG\vW = [\vg_1 \,\,\,\vG_2]$ also has i.i.d.
complex Gaussian entries. Therefore, the elements of $\vg_1$ and
$\vG_2$ are independent. As a result, the quantity $\mct{\vg_1}
(\vG_2\mct{\vG_2})^{-1}\vg_1$ is equivalent to the
signal-to-interference ratio (SIR) of a $\Ne$-branch minimum mean
square error (MMSE) diversity combiner with $\Na-1$ interferers. The
complementary cumulative distribution function of $X=\mct{\vg_1}
(\vG_2\mct{\vG_2})^{-1}\vg_1$ is given in~\cite{gao_98} as
\vspace{-1mm}
\begin{eqnarray}\label{eq:}
    R_X(x)  &=&
    \frac{\sum_{k=0}^{\Ne-1}\binom{\Na-1}{k}x^k}{(1+x)^{\Na-1}}.
\end{eqnarray}
Therefore, we can rewrite (\ref{eq:Cap2}) in an integral form as
\begin{eqnarray}
    \!\!\!C_2 \!\!\!\! &=&\!\!\!\!\! E_{\vh}\Big\{\int_0^{\infty} \log_2\Big(1+\frac{\Na-1}{z-1} x\Big) f_X(x)
    \mathrm{d}x\Big\} \nonumber\\
    &=&\!\!\!\!\! E_{\vh}\Big\{\frac{1}{\ln2}\! \int_0^{\infty}
    \frac{\Na\!-\!1}{z\!-\!1}\Big(1\!+\!\frac{\Na\!-\!1}{z\!-\!1}x\Big)^{-1}\!R_X(x)\mathrm{d}x\Big\}\label{eq:IntUseCCDF}\\
    &=&\!\!\!\!\!E_{\vh}\Big\{\frac{1}{\ln2} \sum_{k=0}^{\Ne-1}\binom{\Na\!-\!1}{k}\nonumber\\
    &&\,\,\,\,\,\,\,\,\,\,\,\,\times\int_0^{\infty}\Big(\frac{z-1}{\Na-1}+x\Big)^{-1}
    (1+x)^{1-\Na}x^k\mathrm{d}x\Big\}\nonumber\\
    &=&\!\!\!\!\!E_{\vh}\Big\{\frac{1}{\ln2}\sum_{k=0}^{\Ne-1}\!\binom{\Na\!-\!1}{k}\frac{\Na\!-\!1}{z\!-\!1}
    \text{B}(k\!+\!1,\Na\!-\!1\!-\!k)\nonumber\\
    &&\,\,\,\,\,\,\,\,\,\,\,\,\times\,_2\text{F}_1\Big(1,k\!+\!1;\Na;\frac{z\!-\!\Na}{z\!-\!1}\Big)\Big\},\label{eq:Cap2C}
\end{eqnarray}
where $f_X(x)$ denotes the probability density function of $X$,
$\text{B}(\alpha,\beta) =
\frac{\Gamma(\alpha)\Gamma(\beta)}{\Gamma(\alpha+\beta)}$ is the
Beta function and $_2\text{F}_1(\cdot)$ is the Gauss hypergeometric
function. Note that (\ref{eq:IntUseCCDF}) is obtained using
integration by parts, and (\ref{eq:Cap2C}) is obtained using an
integration identity given in~\cite{gradshteyn_07}.

After deriving expressions for $C_1$ and $C_2$, a lower bound on the
ergodic secrecy capacity can now be obtained as $C = [C_1 - C_2]^+$,
where $[\alpha]^+=\max\{0,\alpha\}$. This is a data rate that can be
always guaranteed for the secure communication (without knowing the
noise level at Eves). For systems with adaptive power allocation,
the ergodic secrecy capacity lower bound is given as
\begin{eqnarray}\label{eq:AveCapAdapt}
    \!\!\!\!\!\!\!C \!\!\!\!\!&=& \!\!\!\!\!\frac{1}{\ln2}\Big[E_{\vh} \Big\{\ln \Big(1 \!+\! \frac{P}{z} \|\vh\|^2\Big)
    -\!\sum_{k=0}^{\Ne-1}\binom{\Na\!-\!1}{k}\frac{\Na\!-\!1}{z\!-\!1}\nonumber\\
    \!\!\!\!\!\!\!&& \!\!\!\!\! \times\text{B}(k\!+\!1,\Na\!-\!1\!-\!k)\,_2\text{F}_1\Big(1,k\!+\!1;\Na;\frac{z\!-\!\Na}{z\!-\!1}\Big)\Big\}\Big]^+
\end{eqnarray}
where {\em{$z$ is a function of $\vh$}}. For systems with
non-adaptive power allocation, the ergodic secrecy capacity lower
bound is given as
\begin{eqnarray}\label{eq:AveCap}
    \!\!\!\!\!\!\!C \!\!\!\!\!&=& \!\!\!\!\!
    \frac{1}{\ln2}\Big[\exp\Big(\frac{z}{P}\Big)\sum_{k=1}^{\Na}\text{E}_{k}\Big(\frac{z}{P}\Big)
   \!-\!\!\sum_{k=0}^{\Ne-1}\binom{\Na\!-\!1}{k}\frac{\Na\!-\!1}{z\!-\!1}\nonumber\\
   \!\!\!\!\!\!\!&& \!\!\!\!\!
    \times\text{B}(k\!+\!1,\Na\!-\!1\!-\!k)\,_2\text{F}_1\Big(1,k\!+\!1;\Na;\frac{z\!-\!\Na}{z\!-\!1}\Big)\Big]^+
\end{eqnarray}
where {\em{$z$ is a constant independent of $\vh$}}.

\begin{figure}[!t]
\centering\vspace{-3mm}
\includegraphics[width=1.05\columnwidth]{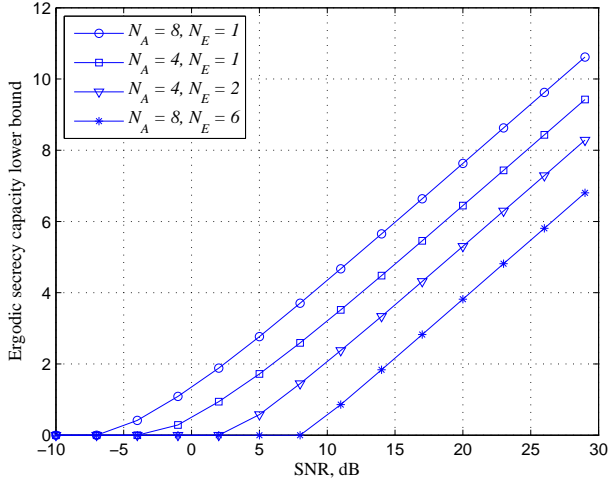}
\vspace{-3mm} \caption{Ergodic secrecy capacity lower bound $C$ in
(\ref{eq:AveCap}) versus SNR $P$ for systems with different numbers
of antennas. The ratio of power allocation is set to $\phi = 0.5$.}
\label{fig:MISOMEcapMonteAndAnalytic}
\end{figure}

Fig.~\ref{fig:MISOMEcapMonteAndAnalytic} shows the ergodic secrecy
capacity lower bound $C$ in (\ref{eq:AveCap}) for systems with
different numbers of antennas. We see that the presence of multiple
colluding Eves dramatically reduces the secrecy rate, compared with
the case of non-colluding Eves. Furthermore, the secrecy rate
quickly reduces to zero at low to moderate SNR.

In the following subsections, we aim to give simplified or
approximated expressions of the secrecy capacity lower bound in two
special scenarios. These expressions will be used to obtain
analytical results and useful insights on the optimal power
allocation in Section~\ref{sec:PowAll}. Note that the derived
approximation may not be an achievable secrecy rate, although it is
useful for the design of power allocation.

\subsection{Non-colluding Eavesdroppers} \label{sec:}

In the case where Eves cannot collude, we have $\Ne=1$.
Then $C_2$ in (\ref{eq:Cap2C}) reduces to
\begin{eqnarray}\label{eq:Cap2CNonCollude}
    \!\!\!\!\!\!C_2 \!\!\!&=&\!\!\! E_{\vh}\Big\{\frac{1}{\ln2}\frac{1}{z-1}
    \,_2\text{F}_1\Big(1,1;\Na;\frac{z-\Na}{z-1}\Big)\Big\}\nonumber\\
    &=& \!\!\! E_{\vh}\Biggl\{\frac{1}{\ln2}\Big(\frac{\Na-1}{\Na-z}\Big)^{\Na-1}\nonumber\\
    &&\,\,\,\,\,\,\,\,\,\times\left(\ln\Big(\frac{\Na\!-\!1}{z\!-\!1}\Big)
    \!-\!\!\sum_{l=1}^{\Na-2}\frac{1}{l}
    \Big(\frac{\Na\!-\!z}{\Na\!-\!1}\Big)^l\right)\Biggr\},
\end{eqnarray}
where (\ref{eq:Cap2CNonCollude}) is obtained using an identity for
the Gauss hypergeometric function derived in Appendix~\ref{app:B}.
This can then be substituted into $C = [C_1 - C_2]^+$ to yield
simplified expressions for the ergodic secrecy capacity lower bound.

\subsection{Large $\Na$ Analysis} \label{sec:}

$C_1$ in (\ref{eq:AveCap1}) can be rewritten as
\begin{eqnarray}
    C_1 \!\!\!&=&\!\!\!  E_{\vh}\Big\{ \log_2 \Big(1 + \frac{P}{z} \|\vh\|^2\Big)
    \Big\}\nonumber\\
    &=&\!\!\! \log_2 \Na + E_{\vh}\Big\{ \log_2 \Big(\frac{1}{\Na}
    + \frac{P}{z} \frac{\|\vh\|^2}{\Na}\Big)\Big\}.
\end{eqnarray}
The law of large numbers implies that $\lim_{\Na \to \infty}
\|\vh\|^2/\Na = 1$. Hence we focus on the non-adaptive power
allocation strategy where $z$ is a constant. In the large $\Na$
limit, we have
\begin{eqnarray}
    \lim_{\Na\rightarrow\infty} (C_1 \!-\! \log_2 \Na)
    \!\!\!\!&=&\!\!\!\!\!  \lim_{\Na\rightarrow\infty}\!
    E_{\vh}\Big\{ \log_2 \Big(\frac{1}{\Na}
    \!+\! \frac{P}{z} \frac{\|\vh\|^2}{\Na}\Big)\Big\}\nonumber\\
    \!\!\!&=&\!\!\! \log_2 \frac{P}{z}.
\end{eqnarray}
That is to say, the difference between $C_1$ and $\log_2\Na$
approaches $\log_2 \frac{P}{z}$ as $\Na$ increases. Therefore, in
the large $\Na$ regime, we have\footnote{The notation $f(x) =
o(g(x))$ implies that $\lim_{x \rightarrow \infty}
\frac{f(x)}{g(x)}=0$. This limit is taken \wrt $\Na$ in
(\ref{eq:InstCapLarge}).}
\begin{eqnarray}\label{eq:InstCapLarge}
    C_1 &=&  \log_2 \Big(\frac{\Na P}{z} \Big) + o(1).
\end{eqnarray}

From the law of large numbers, we also know that $\lim_{\Na \to
\infty} \vG_2\mct{\vG_2}/(\Na-1) = \vI$. Using (\ref{eq:Cap2}) with
the non-adaptive power allocation strategy, we have
\begin{eqnarray}\label{eq:Cap2CLarge}
    \!\lim_{\Na\rightarrow\infty}\!\!\! C_2 \!\!\!\!\!&=&\!\!\!\!\!\! \lim_{\Na\rightarrow\infty}\!\!
    E_{\vg_1, \vG_2}\Big\{\!\log_2\!\Big(1\!+\!\frac{1}{z\!-\!1}
    \mct{\vg_1} \Big(\frac{\vG_2\mct{\vG_2}}{\Na\!-\!1}\Big)^{-1}\!\!\vg_1\!\Big)\!\Big\}\nonumber\\
    &=&\!\!\! E_{\vg_1} \Big\{ \log_2 \Big(1 + \frac{1}{z-1}\|\vg_1\|^2\Big) \Big\}\nonumber\\
    &=&\!\!\! \frac{1}{\ln2}\exp(z-1)
    \sum_{k=1}^{\Ne}\text{E}_{k}(z-1),
\end{eqnarray}
where $\|\vg_1\|^2$ has a Gamma distribution with parameters
$(\Ne,1)$. We see from (\ref{eq:Cap2CLarge}) that altering the
number of antennas at Alice does not affect the channel capacity
between Alice and Eves in the large $\Na$ limit.

The ergodic secrecy capacity lower bound in the large $\Na$ regime
is then given by
\begin{eqnarray}\label{eq:CapLarge}
    C \!=\! \frac{1}{\ln 2} \Big[ \ln\! \Big(\frac{\Na P}{z}\Big)
    \!-\!\exp(z\!-\!1)\! \sum_{k=1}^{\Ne}\text{E}_{k}(z\!-\!1)\! +\! o(1)\Big]^+.
\end{eqnarray}
In Section~\ref{sec:PowAll}, we will use the expression (dropping
$o(1)$) in (\ref{eq:CapLarge}) as an approximation of the secrecy
capacity lower bound for systems with large $\Na$ to study the
optimal power allocation.

\begin{figure}[!t]
\centering\vspace{-3mm}
\includegraphics[width=1.05\columnwidth]{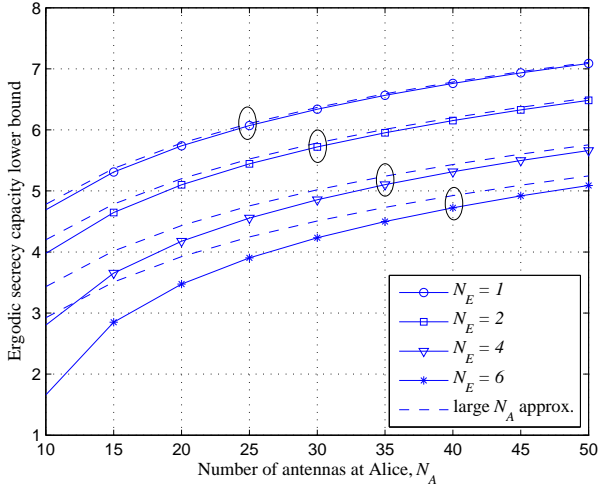}
\vspace{-3mm} \caption{Ergodic secrecy capacity lower bound $C$ in
(\ref{eq:AveCap}) at 10 dB versus the number of antennas at Alice
$\Na$ for systems with different numbers of colluding eavesdroppers.
The large $\Na$ approximations of $C$ in (\ref{eq:CapLarge}) are
shown as the dashed lines. The ratio of power allocation is set to
$\phi = 0.5$.} \label{fig:MISOMEcapLargeAnt}
\end{figure}

Fig.~\ref{fig:MISOMEcapLargeAnt} shows the ergodic secrecy capacity
lower bound $C$ in (\ref{eq:AveCap}) as well as its large $N_A$
approximation in (\ref{eq:CapLarge}). We see that (\ref{eq:AveCap})
converges to (\ref{eq:CapLarge}) as $N_A$ increases. The convergence
is fast for small number of colluding Eves, e.g. $N_E = 2$, and is
slow for large number of colluding Eves, e.g. $N_E = 6$.

\section{Optimal Power Allocation} \label{sec:PowAll}

In this section, we study the optimal power allocation between the
information bearing signal and the artificial noise. As we have
discussed, the power allocation strategy can be either adaptive or
non-adaptive. The former depends on every realization of the channel
gain while the latter is fixed for all channel realizations. The
objective function for this optimization problem is the ergodic
secrecy capacity lower bound. The closed-form expressions derived in
the previous section greatly reduce the computational complexity of
the optimization process. In the following, we first study the case
of non-colluding eavesdroppers and then look at the case of
colluding eavesdroppers.

\subsection{Non-colluding Eavesdropper Case}

The optimal value of $\phi$ or $z$ can be easily found numerically
using the capacity lower bound expressions derived in
Section~\ref{sec:SecCap}. Moveover, these expressions enable us to
analytically obtain useful insights into the optimal $z$ in the high
SNR regime as follows.

In the high SNR regime, \ie $P \gg 1$, $C_1$ in (\ref{eq:AveCap1})
can be approximated as
\begin{eqnarray}\label{eq:C1highSNRapprox}
   C_1 &\approx& E_{\vh} \Big\{\log_2\Big(\frac{P}{z} \|\vh\|^2\Big)
   \Big\}\nonumber\\
   &=& E_{\vh} \{ \log_2(P\|\vh\|^2)\} - E_{\vh}\{\log_2 z\}.
\end{eqnarray}
We see in (\ref{eq:C1highSNRapprox}) that $E_{\vh} \{
\log_2(P\|\vh\|^2)\}$ is a constant and $E_{\vh}\{\log_2 z\}$ does
not directly depend on $\vh$ although $z$ may be a function of
$\vh$. Therefore, the high SNR approximation of the secrecy capacity
lower bound does not have $\vh$ in its expression (except for the
expectation over $\vh$). Consequently, for any value of $\vh$, the
optimal $z$ that maximizes the high SNR approximation of the secrecy
capacity lower bound is the same. In other words, the value of $\vh$
is irrelevant in finding the optimal power allocation. Therefore,
the adaptive power allocation strategy does not need to be
considered at high SNR.

The optimal value of $z$ in the high SNR regime satisfies
\begin{eqnarray}\label{eq:dC}
\frac{\mathrm{d}C}{\mathrm{d}z}= \frac{\mathrm{d}C_1}{\mathrm{d}z}-\frac{\mathrm{d}C_2}{\mathrm{d}z}
=-\frac{1}{z\ln2} -\frac{\mathrm{d}C_2}{\mathrm{d}z} = 0,
\end{eqnarray}
where the derivative of $C_2$ \wrt $z$
can be computed in closed-form using (\ref{eq:Cap2CNonCollude}).

In the special case of $\Na = 2$, (\ref{eq:dC}) is reduces to
\begin{eqnarray}\label{eq:dCNa2}
-\frac{1}{z}- \frac{1}{(z-2)(z-1)} + \frac{\ln(z-1)}{(z-2)^2} = 0.
\end{eqnarray}
The solution to the above equation is given by $z = 2$. It can be
shown that $\lim_{z\to2}\frac{\mathrm{d}^2
C}{\mathrm{d}z^2} < 0$. Hence the optimal ratio of power allocation
is given by $\phi=0.5$, that is to say, equal power allocation
between the information signal and the artificial noise is the
optimal strategy in the high SNR regime for $\Na = 2$.

For large $\Na$, using (\ref{eq:Cap2CLarge}) with $\Ne = 1$, we have
\begin{eqnarray}\label{eq:dC2Large}
 \frac{\mathrm{d}C_2}{\mathrm{d}z}\!\!\! &=& \!\!\!\frac{1}{\ln2} \Big(\exp(z\!-\!1)\text{E}_{1}(z\!-\!1)
 \!-\!\exp(z\!-\!1)\text{E}_{0}(z\!-\!1) \Big)\nonumber\\
    &=&\!\!\! \frac{1}{\ln2} \Big(\exp(z-1)\text{E}_{1}(z-1)
 -(z-1)^{-1} \Big).
\end{eqnarray}
Hence the optimal value of $z$ satisfies
\begin{eqnarray}\label{eq:}
-\frac{1}{z}- e^{z-1}\text{E}_{1}(z-1)+\frac{1}{z-1} = 0,
\end{eqnarray}
which gives $z = 1.80$. It can be shown that at
$z=1.80$, $\frac{\mathrm{d}^2 C}{\mathrm{d}z^2} < 0$.
Hence the optimal ratio of power allocation is given by $\phi=0.55$ in
the high SNR regime for sufficiently large $\Na$. We see that the
difference between the optimal values of $\phi$ for the smallest $\Na$
(\ie $\Na = 2$) and asymptotically large $\Na$ is very small.

\begin{figure}[!t]
\centering\vspace{-3mm}
\includegraphics[width=1.05\columnwidth]{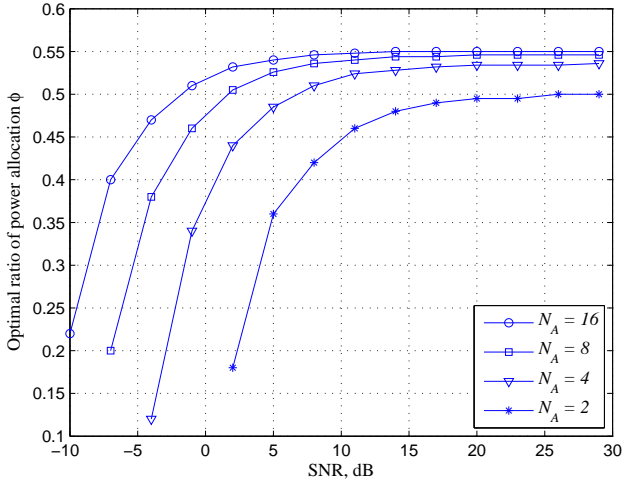}
\vspace{-3mm} \caption{Optimal ratio of power allocation $\phi$
versus SNR $P$ for different numbers of antennas at Alice $\Na$. The
non-adaptive power allocation strategy is used. The values of $\phi$
are shown for SNRs at which the ergodic secrecy capacity lower bound
is positive.} \label{fig:optPower}
\end{figure}

Fig.~\ref{fig:optPower} shows the optimal values of $\phi$ using the
non-adaptive power allocation strategy for systems with different
numbers of antennas at Alice $\Na$. The values of $\phi$ are shown
for SNRs at which the ergodic secrecy capacity lower bound is
positive. The general trend is that more power needs to be allocated
to the information signal as SNR or $\Na$ increases. In the high SNR
regime, we see that the optimal values of $\phi$ converge to
constant values. For $\Na=2$, the optimal value of $\phi$ converges
to 0.5, which agrees with our analytical result. Furthermore, this
constant value only increases slightly with $\Na$, and the maximum
value is 0.55 which agrees with our large $\Na$ analysis. These
observations suggest that a near-optimal yet simple power allocation
strategy at moderate to high SNR values is the equal power
allocation between the information signal and the artificial noise.

\begin{figure}[!t]
\centering\vspace{-3mm}
\includegraphics[width=1.05\columnwidth]{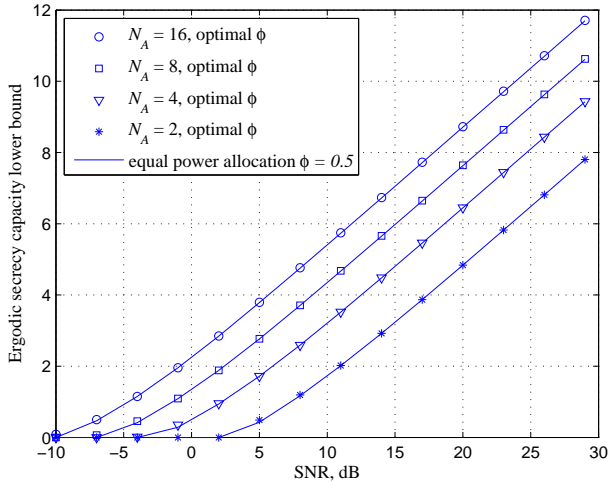}
\vspace{-3mm} \caption{Ergodic secrecy capacity lower bound $C$ in
(\ref{eq:AveCap}) versus SNR $P$ for different numbers of antennas
at Alice $\Na$. The non-adaptive power allocation strategy is used.
The ergodic secrecy capacity lower bound with equal power allocation
for each case, indicated by the solid line, is also shown for
comparison.} \label{fig:capWithOptAndEquPower}
\end{figure}

Fig.~\ref{fig:capWithOptAndEquPower} shows the ergodic secrecy
capacity lower bound $C$ in (\ref{eq:AveCap}) with the optimized
$\phi$ using the non-adaptive power allocation strategy. For
comparison, we also include the capacity lower bound with equal
power allocation, \ie $\phi = 0.5$, indicated by the solid lines. We
see that the equal power allocation strategy achieves nearly the
same secrecy rate as the optimal non-adaptive power allocation in
all cases over a wide range of SNR values. This confirms that equal
power allocation is a simple and generic strategy which yields close
to optimal performance in terms of the derived achievable secrecy
rate.

\begin{figure}[!t]
\centering\vspace{-3mm}
\includegraphics[width=1.05\columnwidth]{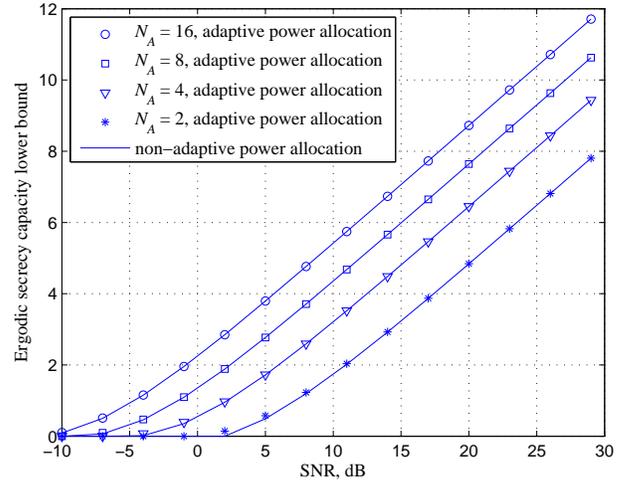}
\vspace{-3mm} \caption{Ergodic secrecy capacity lower bound $C$ in
(\ref{eq:AveCapAdapt}) and (\ref{eq:AveCap}) versus SNR $P$ for
different numbers of antennas at Alice $\Na$. Both the adaptive and
non-adaptive power allocation strategies are used, indicated by the
markers and the lines, respectively.}
\label{fig:capWithAdaptivePowerSplit}
\end{figure}

Fig.~\ref{fig:capWithAdaptivePowerSplit} shows the ergodic secrecy
capacity lower bound $C$ in (\ref{eq:AveCapAdapt}) and
(\ref{eq:AveCap}) with the optimized $\phi$, using both the adaptive
and non-adaptive power allocation strategies. For the adaptive power
allocation, we apply a linear search on $\phi$ to find the optimal
value that maximizes the secrecy capacity lower bound for each
realization of $\vh$. The maximum value of the secrecy capacity
lower bound for each channel realization is recorded and the ergodic
secrecy capacity lower bound is then computed using the distribution
of $\vh$. We see that there is no difference between the secrecy
rate achieved by the adaptive and non-adaptive strategies over a
wide range of SNR values. The adaptive strategy only gives marginal
advantage when the secrecy rate is close to zero. This result
suggests that the non-adaptive power allocation strategy is
sufficient to achieve almost the best possible secrecy rate
performance.\footnote{The same result is found for the colluding
Eves case. The numerical results are omitted for brevity.} For this
reason, we will only focus on the non-adaptive scheme in the rest of
this paper.

\subsection{Colluding Eavesdropper Case}

As we have seen in Fig.~\ref{fig:MISOMEcapMonteAndAnalytic}, the
presence of multiple colluding Eves severely degrades the secrecy
rate. Therefore, it is essential for Alice to have a relatively
large number of antennas to maintain a good secure communication
link. For any value of $\Ne$, the optimal value of $\phi$ or $z$ can
be easily found numerically using the closed-form capacity lower
bound expression given in Section~\ref{sec:SecCap}. As the number of
antennas at Alice is desired to be large, we carry out large $\Na$
analysis to obtain an asymptotic result on optimal $z$ in the high
SNR regime as follows.

In the high SNR regime with large $\Na$, $C$ in
(\ref{eq:CapLarge}) can be approximated as
\begin{eqnarray}\label{eq:CapLargeHighSNR}
    C \approx \frac{1}{\ln 2} \Big[ \ln (\Na P) - \ln z
    -\exp(z-1) \sum_{k=1}^{\Ne}\text{E}_{k}(z-1)\Big].
\end{eqnarray}
By taking the derivative of $C$ \wrt $z$, the optimal $z$ satisfies
\begin{eqnarray}\label{eq:dClargeNa}
    -\frac{1}{z} - e^{z-1} \text{E}_{\Ne}(z-1)+\frac{1}{z-1} = 0.
\end{eqnarray}
Using $e^{z-1} \text{E}_{\Ne}(z-1) \approx (z-1+\Ne)^{-1}$ from~\cite{abramowitz_70},
which is accurate when either $\Ne$ or $z$ is large, (\ref{eq:dClargeNa}) reduces to
\begin{eqnarray}\label{eq:dClargeNa2}
    -\frac{1}{z} - \frac{1}{z-1+\Ne}+\frac{1}{z-1} = 0.
\end{eqnarray}
Hence the optimal $z$ is given by
\begin{eqnarray}\label{eq:OptimalPowerHighLargeDouble}
    z^* = 1+\sqrt{\Ne}.
\end{eqnarray}

From (\ref{eq:OptimalPowerHighLargeDouble}) we see that the optimal
value of $z$ only depends on $\Ne$ in the high SNR and large
antenna regime. Moreover,
(\ref{eq:OptimalPowerHighLargeDouble}) suggests that more power should
be used to generate artificial noise when the number of Eves increases.

\begin{figure}[!t]
\centering\vspace{-3mm}
\includegraphics[width=1.05\columnwidth]{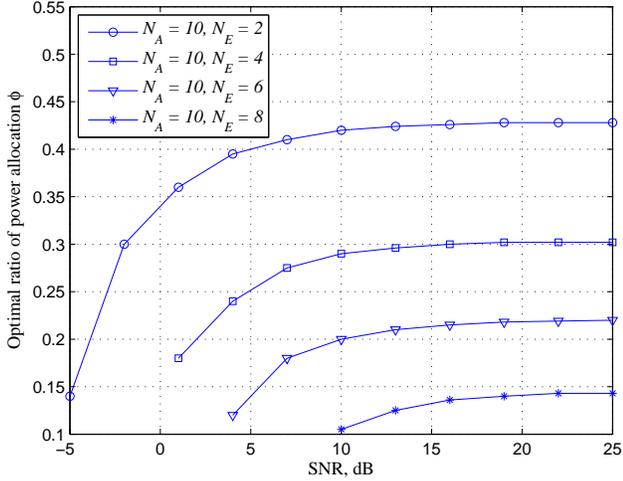}
\vspace{-3mm} \caption{Optimal ratio of power allocation $\phi$
versus SNR $P$ for systems with different numbers of colluding Eves
$\Ne$. The values of $\phi$ are shown for SNRs at which the ergodic
secrecy capacity lower bound is positive.}
\label{fig:MISOMEoptPower}
\end{figure}

Fig.~\ref{fig:MISOMEoptPower} shows the optimal value of $\phi$ for
systems with different numbers of colluding Eves $\Ne$.
Similar to the non-colluding Eves case, we see that more power should be used to
transmit the information signal as SNR increases. The optimal value of $\phi$ stays constant in the high SNR
regime. Furthermore, the optimal value of $\phi$ for colluding Eves case
is usually much smaller than 0.5, \ie equal power allocation, which is near
optimal for non-colluding Eves case. In particular, the optimal
$\phi$ reduces as $\Ne$ grows, which implies that more power
should be allocated to generate the artificial noise as the number of colluding
Eves increases. This observation agrees with our analytical insight and intuition.

\begin{figure}[!t]
\centering\vspace{-3mm}
\includegraphics[width=1.05\columnwidth]{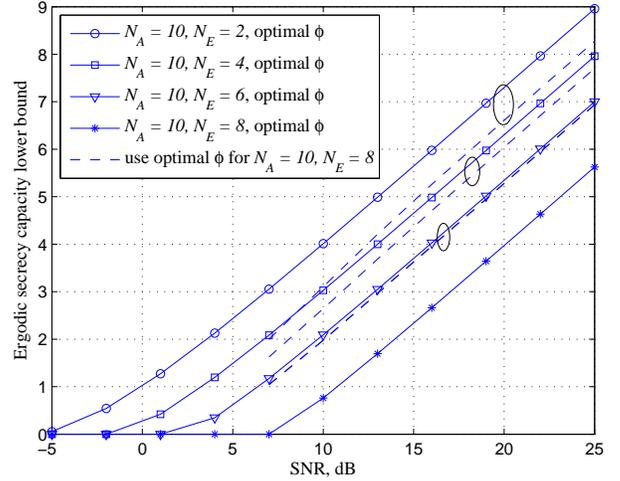}
\vspace{-3mm} \caption{Ergodic secrecy capacity lower bound $C$ in
(\ref{eq:AveCap}) versus SNR $P$ for systems with different numbers
of colluding Eves $\Ne$. The solid lines with markers indicate $C$
achieved with optimal values of $\phi$ for the corresponding system.
The dashed lines indicate $C$ achieved with value of $\phi$
optimized for $\Ne = 8$, which represents the case where the power
allocation was initially designed for $\Ne = 8$, but the current
value of $\Ne$ reduces from 8 and the power allocation is not
redesigned.} \label{fig:MISOMEcapWithoptPower}
\end{figure}

Fig.~\ref{fig:MISOMEcapWithoptPower} shows the ergodic secrecy
capacity lower bound $C$ in (\ref{eq:AveCap}) for systems with
different $\Ne$. Here, we investigate the sensitivity in the secrecy
rate to the design of power allocation. Consider a scenario where
the total number of Eves that can collude is 8, and hence Alice has
optimized $\phi$ for $\Ne = 8$. When $\Ne$ changes, the power
allocation parameter $\phi$ does not need to be optimized again as
long as $\Ne$ stays reasonably close to 8, \eg $\Ne=6$, since the
value of $\phi$ optimized for $\Ne=8$ still works well for $\Ne=6$
(with a power loss of 0.2 dB) as shown in
Fig.~\ref{fig:MISOMEcapWithoptPower}. However, redesigning of $\phi$
becomes important when $\Ne$ is considerably different from~8, \eg
$\Ne = 2$ to 4. For example, if $\Ne$ changes from 8 to 4, a power
loss of approximately 1~dB will incur if Alice still uses the value
of $\phi$ optimized for $\Ne = 8$, as shown in
Fig.~\ref{fig:MISOMEcapWithoptPower}.

\begin{figure}[!t]
\centering\vspace{-3mm}
\includegraphics[width=1.05\columnwidth]{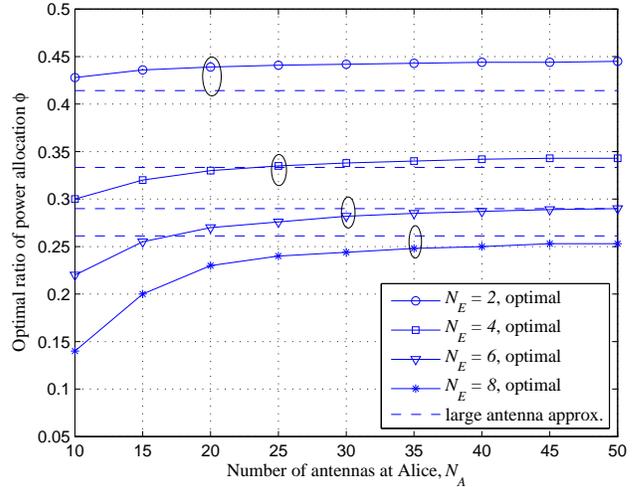}
\vspace{-3mm} \caption{The ratio of power allocation $\phi$ at 20 dB
versus the number of antennas at Alice $\Na$ for systems
with different numbers of colluding Eves $\Ne$. The solid
lines with markers indicate the optimal values of $\phi$, while the dashed lines
indicate the values of $\phi$ from the large antenna approximation given in
(\ref{eq:OptimalPowerHighLargeDouble}).}
\label{fig:MISOMEoptPowerLargeAnt}
\end{figure}

\begin{table*}[t]
\setcounter{mytempeqncnt}{\value{equation}}
\setcounter{equation}{32}
\normalsize
    \begin{eqnarray}\label{eq:CriSNR}
   P_C < z\left[\frac{\Na}{\sum_{k=0}^{\Ne-1}\!\binom{\Na-1}{k}\frac{\Na-1}{z-1}
    \text{B}(k\!+\!1,\Na\!-\!1\!-\!k)\,_2\text{F}_1\Big(1,k\!+\!1;\Na;\frac{z-\Na}{z-1}\Big)}
     -\frac{\Na\!+\!1}{2}\right]^{-1}.
\end{eqnarray}
\hrulefill
\begin{eqnarray}\label{eq:CriSNRnonCollude}
   P_C < z\left[\frac{\Na}{\Big(\frac{\Na-1}{\Na-z}\Big)^{\Na-1}
    \left(\ln\Big(\frac{\Na-1}{z-1}\Big)-\sum_{l=1}^{\Na-2}\frac{1}{l}\Big(\frac{\Na-z}{\Na-1}\Big)^l
    \right)} -\frac{\Na+1}{2}\right]^{-1}.
\end{eqnarray}
\setcounter{equation}{\value{mytempeqncnt}}
     \hrulefill
 \end{table*}

We also provide numerical verification of the optimal power allocation
obtained from the large antenna approximation in the high SNR regime.
Fig.~\ref{fig:MISOMEoptPowerLargeAnt} shows the ratio of power
allocation $\phi$ at 20~dB versus the number of antennas at Alice $\Na$
for systems with different numbers of colluding
Eves $\Ne$. For a fixed $\Ne$, we see that the optimal
value of $\phi$ increases with $\Na$ and reaches a constant value
when $\Na$ is sufficiently large. This agrees with our analytical insight
that the optimal power allocation depends on $\Ne$ but not on $\Na$ when
$\Na$ is large. The asymptotic constant value of $\phi$ is close to
the analytical value given in (\ref{eq:OptimalPowerHighLargeDouble})
obtained from the large antenna approximation.

In the system model, we have assumed fixed power transmission over
time. When variable power transmission is allowed subject to an
average power constraint, the achievable secrecy rate can be
increased by having temporal power allocation according to the
channel gain at each time instant. From the derived secrecy rate
expression, we see that the transmit power only affects the
transmission rate between Alice and Bob. The existing study on the
point-to-point channel capacity, e.g. in~\cite{yoo_goldsmith_06},
showed that the temporal power optimization gives little capacity
gain provided that the spatial power optimization is used.

In reality, the noise is always present at the eavesdroppers and
hence, the designed power allocation strategy is not the optimal
strategy in practice. If the eavesdroppers' noise levels are known
to the transmitter and hence are taken into account in the secrecy
rate expression, the efficiency of using artificial noise in
degrading the capacity between Alice and Eve is reduced. Therefore,
more power should be used to transmit the information signal.

\section{Critical SNR for Secure Communications}\label{sec:CriSNR}

Another important aspect of secure communication is the minimum SNR
required for a positive secrecy rate, which is a critical parameter
in wideband communications. With the closed-form expression of the
secrecy capacity lower bound derived in Section~\ref{sec:SecCap},
one can numerically find the critical SNR with low computational
complexity. In this section, we derive a closed-form upper bound on
the critical SNR which is useful in the design of wideband
communications.

Using properties of the exponential integral function
in~\cite{abramowitz_70}, (\ref{eq:AveCap1C}) can be bounded from
below as
\begin{eqnarray}\label{eq:C1lowSNR1}
   C_1 &>& \frac{1}{\ln 2} \sum_{k=1}^{\Na} \frac{1}{\frac{z}{P}+k},
\end{eqnarray}
which is asymptotically tight as the SNR approaches zero, \ie $P
\rightarrow 0$. Using the convexity of (\ref{eq:C1lowSNR1}) in $k$,
we can further bound $C_1$ as
\begin{eqnarray}\label{eq:C1lowSNR2}
   C_1 &>& \frac{1}{\ln 2} \frac{\Na}{\frac{z}{P}+\frac{\Na+1}{2}},
\end{eqnarray}
which is also asymptotically tight as the SNR approaches zero. Using
the lower bound on $C_1$ in (\ref{eq:C1lowSNR2}) and $C_2$ in
(\ref{eq:Cap2C}), the ergodic secrecy capacity lower bound can be
further bounded from below as
\begin{eqnarray}\label{eq:ClowSNR}
  \!\!\!\!\!\!\!\!\!\!C \!\!\!\!&>&\!\!\!\! \frac{1}{\ln 2} \frac{\Na}{\frac{z}{P}+\frac{\Na+1}{2}}
       -\frac{1}{\ln2}\sum_{k=0}^{\Ne-1}\!\binom{\Na\!-\!1}{k}\frac{\Na\!-\!1}{z\!-\!1}\nonumber\\
    \!\!\!\!\!\!\!\!\!\!&&\times\text{B}(k\!+\!1,\Na\!-\!1\!-\!k)\,_2\text{F}_1\Big(1,k\!+\!1;\Na;\frac{z\!-\!\Na}{z\!-\!1}\Big).
\end{eqnarray}

The critical SNR, denoted by $P_C$, is the SNR at which $C$ drops to
zero. With the lower bound on $C$ given in (\ref{eq:ClowSNR}), an
upper bound on $P_C$ can be found as (\ref{eq:CriSNR}) on the top of the page. In the case of non-colluding eavesdroppers, \ie $\Ne=1$,
(\ref{eq:CriSNR}) reduces to (\ref{eq:CriSNRnonCollude}) on the top of the page. The upper bound in (\ref{eq:CriSNR}) or (\ref{eq:CriSNRnonCollude})
indicates a minimum SNR that guarantees a positive secrecy rate.
Since (\ref{eq:CriSNR}) and (\ref{eq:CriSNRnonCollude}) are
asymptotically tight at low SNR, they can be used to fine tune the
power allocation parameter $z$ to minimize $P_C$.

\addtocounter{equation}{2}

\begin{figure}[!t]
\centering\vspace{-3mm}
\includegraphics[width=1.05\columnwidth]{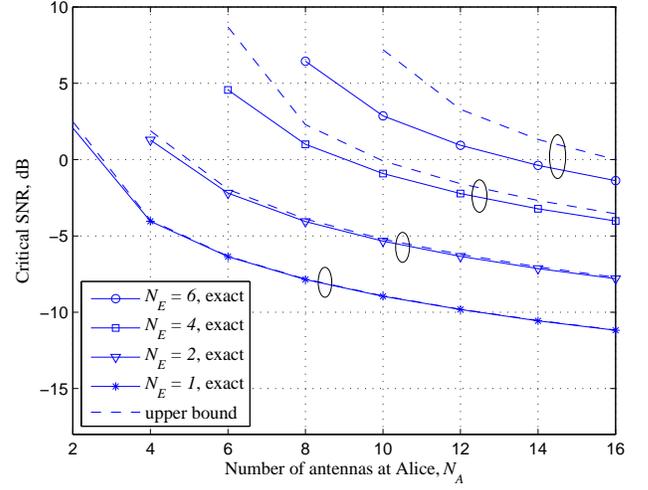}
\vspace{-3mm} \caption{The critical SNR $P_C$ versus number of antennas
at Alice $\Na$ for systems with different numbers of colluding
Eves $\Ne$. The ratio of power allocation is set to $\phi = 0.2$. The
solid lines with markers indicate the exact value of $P_C$, while
the dashed lines indicate the analytical upper bound given in
(\ref{eq:CriSNR}).} \label{fig:MISOMEcriticalSNR} \vspace{-0mm}
\end{figure}

\begin{table*}[t]
\setcounter{mytempeqncnt}{\value{equation}}
\setcounter{equation}{41}
\normalsize

\begin{eqnarray}\label{eq:CriSNRError}
   P_C < \left[\frac{1-\shtilde}{z}\Big(\frac{\Na}{
\sum_{k=0}^{\Ne-1}\binom{\Na-1}{k}\frac{\Na-1}{z-1}
    \text{B}(k+1,\Na-1-k)\,_2\text{F}_1\Big(1,k+1;\Na;\frac{z-\Na}{z-1}\Big)}
     -\frac{\Na+1}{2}\Big)-\shtilde\right]^{-1},
\end{eqnarray}
\setcounter{equation}{\value{mytempeqncnt}}
     \hrulefill
 \end{table*}

Fig.~\ref{fig:MISOMEcriticalSNR} shows the critical SNR $P_C$ versus
number of antennas at Alice $\Na$ for systems with different numbers of
colluding Eves $\Ne$. The power allocation is set to $\phi = 0.2$
in all cases. The general trend is that $P_C$ decreases as $\Na$ increases, and a
higher $P_C$ is required when $\Ne$ increases. These observations
agree with intuition. Furthermore, we see that the analytical upper
bound on $P_C$ is very accurate for the case of non-colluding
Eves. For the case of colluding Eves, the upper bound is reasonably
accurate when $P_C < 0$ dB. The difference between the exact
value of $P_C$ and its upper bound gradually increases as $\Ne$ increases,
which is mainly due to the increase in $P_C$. When $\Ne$ is
relatively large, \eg $\Ne = 6$, one should allocate more power to
generate the artificial noise (\ie reduce $\phi$), as suggested in
Fig.~\ref{fig:MISOMEoptPower}, in order to achieve a lower $P_C$, which
in turn makes the bound tighter.

\section{Effect of Imperfect Channel State Information}\label{sec:ImpCSI}

So far, we have assumed that the CSI can be
perfectly obtained at Alice and Bob. In this section, we investigate the
effect of imperfect CSI by considering channel estimation errors. With imperfect
CSI, the beamforming transmission from Alice to Bob is designed based on
the estimated channel gains rather than the true channel gains. Therefore,
the artificial noise leaks into Bob's channel.

To incorporate imperfect CSI, we consider that Bob performs
the MMSE channel estimation. Therefore, we have
\begin{eqnarray}\label{eq:}
   \vh &=& \hat{\vh} + \tilde{\vh},\\
   \sh &=& \shhat + \shtilde,
\end{eqnarray}
where $\hat{\vh}$ denotes the channel estimate and $\tilde{\vh}$
denotes the estimation error. $\sh$ denotes the variance of each
element in $\vh$. $\shhat$ and $\shtilde$ denote the variance of
each element in $\hat{\vh}$ and $\tilde{\vh}$, respectively. As a
general property of the MMSE estimator for Gaussian
signals~\cite{anderson_moore_79}, $\hat{\vh}$ and $\tilde{\vh}$ are
uncorrelated, each having i.i.d. complex Gaussian entries.

Similar to our system model in Section~\ref{sec:SysMod}, we assume that the
knowledge of $\hat{\vh}$ is available at Alice and Eves.
Therefore, the beamforming vector becomes $\vw_1 =
\mct{\hat{\vh}}/\|\hat{\vh}\|$, and the received symbol at Bob is
given by
\begin{eqnarray}\label{eq:SigModError}
    \!\!y_B \!\!&=&\!\! \hat{\vh}\vx + \tilde{\vh}\vx + n
    = \|\hat{\vh}\| u + \tilde{\vh} \vW [u \,\,\,{\vv}^T]^T + n.
\end{eqnarray}

A capacity lower bound for the channel between Alice and Bob can be
obtained by considering $\tilde{\vh} \vW [u \,\,\,{\vv}^T]^T + n$ as
the worst case Gaussian noise~\cite{hassibi_hochwald_03}. Note that
$\vW$ is a unitary matrix, hence $\tilde{\vh} \vW$ has the same
distribution as $\tilde{\vh}$~\cite{telatar_99}. Therefore, the
ergodic capacity lower bound for the channel between Alice and Bob
is given by
\begin{eqnarray}\label{eq:AveCapError}
    \hat{C}_1 &=& E_{\hat{\vh}} \Big\{\log_2 \Big(1 +
    \frac{\su \|\hat{\vh}\|^2}{\shtilde P + 1}\Big)\Big\}.
\end{eqnarray}

With $\sh$ normalized to unity, we have $\shhat=1-\shtilde$. Since
the elements of $\hat{\vh}$ is i.i.d. complex Gaussian,
$\|\hat{\vh}\|^2$ is a sum of i.i.d. exponential distributed random
variables, which follows a Gamma distribution with parameter $(\Na,
1-\shtilde)$. Therefore, we obtain a closed-form expression for
$\hat{C}_1$ as
\begin{eqnarray}\label{eq:AveCapError2}
    \!\!\!\!\hat{C}_1 \!\!\!\!&=&\!\!\!\!
    \frac{1}{\ln2}\exp\Big(z\frac{\shtilde+P^{-1}}{1-\shtilde}\Big)
    \sum_{k=1}^{\Na}\text{E}_{k}\Big(z\frac{\shtilde+P^{-1}}{1-\shtilde}\Big).
\end{eqnarray}

The presence of channel estimation errors does not affect the signal
reception at Eve given in (\ref{eq:SigMod2}). Therefore, the ergodic
secrecy capacity lower bound can be obtained by subtracting $C_2$
from $\hat{C}_1$ as
\begin{eqnarray}\label{eq:AveCapError3}
    C \!\!\!&=& \!\!\!\frac{1}{\ln2}\Big[\exp\Big(z\frac{\shtilde+P^{-1}}{1-\shtilde}\Big)
    \sum_{k=1}^{\Na}\text{E}_{k}\Big(z\frac{\shtilde+P^{-1}}{1-\shtilde}\Big)\nonumber\\
     \!\!\!&&\!\!\!-\sum_{k=0}^{\Ne-1}\!\binom{\Na\!-\!1}{k}\frac{\Na\!-\!1}{z\!-\!1}
    \text{B}(k\!+\!1,\Na\!-\!1\!-\!k)\nonumber\\
    &&\,\,\,\,\,\,\,\,\,\,\,\,\,\,\,\,\,\,\times\,_2\text{F}_1\Big(1,k\!+\!1;\Na;\frac{z\!-\!\Na}{z\!-\!1}\Big)\Big]^+.
\end{eqnarray}

Following the steps in Section~\ref{sec:CriSNR}, we can also bound $C$ from
below in order to obtain an upper bound on the critical SNR for secure communication
with channel estimation errors as
\begin{eqnarray}\label{eq:C1lowSNR2Error}
   C \!\!\!\!&>&\!\!\!\! \frac{1}{\ln 2} \frac{\Na}{z \frac{\shtilde+P^{-1}}{1-\shtilde}+\frac{\Na+1}{2}}
    \!-\!\frac{1}{\ln2}\!\sum_{k=0}^{\Ne-1}\!\!\binom{\Na\!-\!1}{k}\frac{\Na\!-\!1}{z\!-\!1}\nonumber\\
    \!\!\!\!&&\!\!\!\!\times\text{B}(k\!+\!1,\Na\!-\!1\!-\!k)\,_2\text{F}_1\Big(1,k\!+\!1;\Na;\frac{z\!-\!\Na}{z\!-\!1}\Big).
\end{eqnarray}
And the upper bound on the critical SNR is then given in (\ref{eq:CriSNRError}) on the top of the page,
which is asymptotically tight at low SNR.

\addtocounter{equation}{1}

We now present numerical results on the optimal power allocation
as well as critical SNR in the presence of the channel estimation
errors. For brevity, we focus on the case of non-colluding
eavesdroppers. The trends on the effect of channel estimation errors
observed in the following results also apply to the case of colluding
eavesdroppers.

\begin{figure}[!t]
\centering\vspace{-3mm}
\includegraphics[width=1.05\columnwidth]{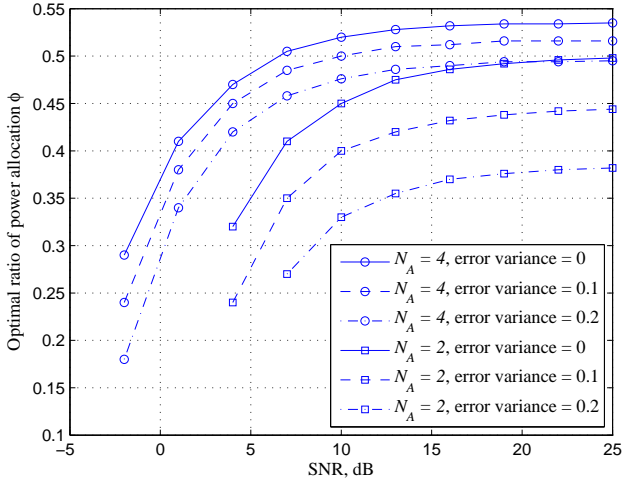}
\vspace{-3mm} \caption{Optimal ratio of power allocation $\phi$
versus SNR $P$ for different numbers of antennas at Alice $\Na$ and
different variances of the channel estimation errors $\shtilde$. The
values of $\phi$ are shown for SNRs at which the ergodic secrecy
capacity lower bound is positive.}
\label{fig:MISOoptPower_withError}
\end{figure}

Fig.~\ref{fig:MISOoptPower_withError} shows the optimal ratio of power
allocation $\phi$ with different channel estimation error
variances $\shtilde$. We see that the channel estimation error has noticeable
impact on the value of $\phi$, especially for small number of antennas
at Alice, \eg $\Na=2$. The general trend is that less power should be allocated to information
signal as channel estimation error increases. This is mainly due to the fact
that the efficiency of improving Bob's signal reception by boosting the transmit power of the information
signal reduces as the channel estimation error increases. On the other hand, the efficiency of
degrading Eve's signal reception by boosting the transmit power of the artificial noise
stays the same regardless of the channel estimation error. Hence, it is
better to create more noise for Eves than to increase the signal
strength for Bob if the CSI is not accurately obtained.

In practical systems, the channel estimation error usually reduces as the
SNR increases, although their exact relationship depends on the training design.
From Fig.~\ref{fig:MISOoptPower_withError}, we can expect that at low to moderate SNR
where the channel estimation error is usually noticeable, the optimal power allocation
is very different from that in the perfect CSI case. While at high SNR where the channel estimation
error is usually small, the optimal power allocation is expected to be very close to
that of the perfect CSI case. Therefore, in practical systems
it is important to take channel estimation error into account when designing the power
allocation at relatively low SNR.

\begin{table}[!t]
\caption{Critical SNR (in dB) for Secure Communications with Equal
Power Allocation}
\label{table:CriSNRdq}%
\centering
\begin{tabular}{c|c||c|c|c|c|c} \hline
     & Error variance & \multicolumn{5}{c}{Number of antennas $\Na$} \\ \cline{3-7}
     & $\shtilde$ & 2 & 4 & 6 & 8 & 10 \\ \hline\hline
     Exact       & 0 & 3.01 & -2.62 & -4.89 & -6.36 & -7.45 \\
     Upper bound & 0 & 6.02 & -1.97 & -4.46 & -6.01 & -7.14 \\ \hline
     Exact       & 0.1 & 4.56 & -1.88 & -4.27 & -5.79 & -6.90 \\
     Upper bound & 0.1 & 9.03 & -1.20 & -3.83 & -5.43 & -6.59 \\ \hline
     Exact       & 0.2 & 6.99 & -1.01 & -3.55 & -5.13 & -6.28 \\
     Upper bound & 0.2 & $\infty$ & -0.26 & -3.08 & -4.76 & -5.96 \\ \hline
\end{tabular}
\end{table}

Table~\ref{table:CriSNRdq} lists the exact values of the critical SNR
$P_C$ as well as the closed-form upper bound given in
(\ref{eq:CriSNRError}) with $\phi=0.5$.  The general trend is that the critical
SNR increases as the channel estimation error increases, which agrees with intuition.
The upper bound gets tighter as $P_C$ reduces (or $\Na$ increases), and is
accurate for $\Na\geq 4$ with an error of less than 1 dB.

\section{Conclusion}

In this paper, we considered the secure communication in the
wireless fading environment in the presence of non-colluding or
colluding eavesdroppers. The transmitter is equipped with multiple
antennas and is able to simultaneously transmit an information
signal to the intended receiver and artificial noise to confuse the
eavesdroppers. We obtained an closed-form expression for the ergodic
secrecy capacity lower bound. We studied the optimal power
allocation between transmission of the information signal and the
artificial noise. In particular, equal power allocation was shown to
be a near optimal strategy in the case of non-colluding
eavesdroppers. When the number of colluding eavesdroppers increases,
more power should be used to generate artificial noise. We also
derived an upper bound on the critical SNR above which the secrecy
rate is positive and this bound was shown to be tight at low SNR.
When imperfect channel state information was considered in the form
of channel estimation errors, we found that it is wise to create
more artificial noise to confuse the eavesdroppers than to increase
the signal strength for the intended receiver. The results obtained
in this work provide various insights into the design and analysis
of secure communication with multi-antenna transmission.

\appendices

\section{Identity for Special Class of Gauss Hypergeometric Function}
\label{app:B}

Here we obtain a simplified expression for the Gauss hypergeometric
function in the form of $_2\text{F}_1(1,1;N+1; x)$ or
$_2\text{F}_1(N,N;N+1; x)$ for integer $N\geq1$. From~\cite{abramowitz_70},
we know that these two forms of the Gauss hypergeometric function are
related to each other by
\begin{eqnarray}\label{eq:2F1relation}
    _2\text{F}_1(1,1;N+1; x) = (1-x)^{N-1}\,_2\text{F}_1(N,N;N+1;
    x).
\end{eqnarray}
Also, we know from~\cite{abramowitz_70} that
\begin{eqnarray}\label{eq:}
    \frac{\mathrm{d}^{N-1}}{\mathrm{d}x^{N-1}}\,_2\text{F}_1(1,1;2; x) \!=\!
    \frac{(1)_{N-1}(1)_{N-1}}{(2)_{N-1}}\,_2\text{F}_1(N,N;N\!+\!1; x),\nonumber
\end{eqnarray}
where $(a)_b$ is the rising factorial. Therefore, we have
\begin{eqnarray}\label{eq:SpecialGauss}
    \!\!\!\!\!\!\!\!&&\!\!\!\!\!\!\!\!\!\!_2\text{F}_1(N,N;N+1; x) \nonumber\\
    \!\!\!\!\!\!\!\!&&\!\!\!\!\!\!\!\!\!\!= \frac{(2)_{N-1}}{(1)_{N-1}(1)_{N-1}}
    \frac{\mathrm{d}^{N-1}}{\mathrm{d}x^{N-1}}\,_2\text{F}_1(1,1;2;
    x)\nonumber\\
    \!\!\!\!\!\!\!\!&&\!\!\!\!\!\!\!\!\!\!= -\frac{N}{(N\!-\!1)!}\sum_{l=0}^{N-1}\binom{N\!-\!1}{l}
    \frac{\mathrm{d}^{l}}{\mathrm{d}x^{l}}\ln(1\!-\!x)
    \frac{\mathrm{d}^{N-1-l}}{\mathrm{d}x^{N-1-l}}x^{-1},
\end{eqnarray}
where we have used the identity $_2\text{F}_1(1,1;2; x) =
-\ln(1-x)/x$ from~\cite{abramowitz_70}. It is easy to show that
\begin{eqnarray}\label{eq:}
    \frac{\mathrm{d}^{k}}{\mathrm{d}x^{k}}\ln(1-x)
    &=& -\frac{\mathrm{d}^{k-1}}{\mathrm{d}x^{k-1}}(1-x)^{-1}\nonumber\\
    &=& - \frac{(k-1)!}{(1-x)^k},\,\,\,k=1,2,3,...\nonumber\\
    \frac{\mathrm{d}^{k}}{\mathrm{d}x^{k}}z^{-1} &=& \frac{(-1)^k
    k!}{x^{k+1}},\,\,\,k = 0, 1, 2, 3,...\nonumber
\end{eqnarray}
Substituting the above expressions for the derivatives into
(\ref{eq:SpecialGauss}), we obtain an identity expression as
\begin{eqnarray}\label{eq:SpecialGauss2}
    \!\!\!\!\!\!\!\!&&\!\!\!\!\!\!\!\!\!\!_2\text{F}_1(N,N;N+1; x)\nonumber\\
    \!\!\!\!\!\!\!\!&&\!\!\!\!\!\!\!\!\!\!=
    -\frac{N}{(N-1)!}\Big(\ln(1-x)\frac{(-1)^{N-1}(N-1)!}{z^N}\nonumber\\
    &&- \sum_{l=1}^{N-1}
    \frac{(N-1)!}{l!(N-1-l)!}\frac{(l-1)!}{(1-x)^l}\frac{(-1)^{N-1-l}(N-1-l)!}{x^{N-l}}\Big)\nonumber\\
    \!\!\!\!\!\!\!\!&&\!\!\!\!\!\!\!\!\!\!=
    \frac{(-1)^N N}{x^N}\Big(\ln(1-x) - \sum_{l=1}^{N-1}
    \frac{1}{l}\frac{x^l}{(x-1)^l}\Big).
\end{eqnarray}
Using (\ref{eq:2F1relation}), we also have
\begin{eqnarray}\label{eq:SpecialGauss3}
    \!\!\!\!\!\!\!\!&&\!\!\!\!\!\!\!\!\!\!_2\text{F}_1(1,1;N+1; x)\nonumber\\
    \!\!\!\!\!\!\!\!&&\!\!\!\!\!\!\!\!\!\!= \frac{(-1)^N N (1-x)^{N-1}}{x^N}
    \Big(\ln(1\!-\!x) \!-\! \sum_{l=1}^{N-1}
    \frac{1}{l}\frac{x^l}{(x-1)^l}\Big).
\end{eqnarray}

\section*{Acknowledgements}\ThankThree

\bibliographystyle{IEEEtran}
\bibliography{IEEEabrv,journal_v4_double}

\vspace{3mm} \textbf{Xiangyun Zhou} (S'08) received the B.E. (hons.)
degree in electronics and telecommunications engineering from the
Australian National University, Australia, in 2007. He is currently
working toward the Ph.D. degree in engineering and information
technology at the Research School of Information Sciences and
Engineering, the Australian National University. His research
interests are in signal processing for wireless communications,
including MIMO systems, ad hoc networks, relay and cooperative
networks, and physical-layer security.

\vspace{3mm} \textbf{Matthew R.\ McKay} (S'03, M'07) received the combined B.E. degree
in Electrical Engineering and B.IT. degree in Computer Science from
the Queensland University of Technology, Australia, in 2002, and the
Ph.D. degree in Electrical Engineering from the University of
Sydney, Australia, in 2006.  He then worked as a Research Scientist
at the Commonwealth Science and Industrial Research Organization
(CSIRO), Sydney, Australia, prior to joining the faculty at the Hong
Kong University of Science and Technology (HKUST) in 2007, where he
is currently an Assistant Professor.  He is also a member of the
Center for Wireless Information Technology at HKUST.  His research
interests include communications and signal processing; in
particular the analysis and design of MIMO systems, random matrix
theory, information theory, and wireless ad-hoc and sensor networks.

He was awarded a 2006 Best Student Paper Award at IEEE ICASSP'06,
and was jointly awarded the 2006 Best Student Paper Award at IEEE
VTC'06-Spring.  He was also awarded the University Medal upon
graduating from the Queensland University of Technology.

\end{document}